\documentclass[twocolumn,prb,showpacs,preprintnumbers]{revtex4}

\usepackage{graphicx}
\usepackage{epsfig, array, amsmath, delarray, amssymb}

\begin{document}

\newcommand{\rv}{{\mathbf r}}

\preprint{LA-UR-05-0745}

\title{Study of the Pioneer anomaly: A problem set}

\author{Slava G. Turyshev}
\email{turyshev@jpl.nasa.gov}
\affiliation{Jet Propulsion Laboratory, 
California Institute of Technology,
4800 Oak Grove Drive, Pasadena, California 91109}

\author{Michael Martin Nieto}
\email{mmn@lanl.gov}
\affiliation{Theoretical Division, MS-B285, 
Los Alamos National Laboratory,
University of California, Los Alamos, New Mexico 87545}

\author{John D. Anderson}
\email{john.d.anderson@jpl.nasa.gov}
\affiliation{Jet Propulsion Laboratory, 
California Institute of Technology,
4800 Oak Grove Drive, Pasadena, California 91109}


\begin{abstract}
Analysis of the radio-metric tracking data from the Pioneer 10 and 11 spacecraft at distances between 20 and 70 astronomical units from the Sun has consistently indicated the presence of an anomalous, small, and constant Doppler frequency drift. The drift is a blueshift, uniformly changing at the rate of $(5.99 \pm 0.01) \times 10^{-9}$\,Hz/s. The signal also can be interpreted as a constant acceleration of each particular spacecraft of $(8.74 \pm 1.33)\times 10^{-8}$\,cm/s$^2$ directed toward the Sun. This interpretation has become known as the Pioneer anomaly. We provide a problem set based on the detailed investigation of this anomaly, the nature of which remains unexplained.
\end{abstract}

\pacs{04.80.-y, 95.10.Eg, 95.55.Pe}

\maketitle

\section{Introduction}
\label{sec:background}

\begin{figure*}[t!]
\begin{center} 
\epsfig{figure=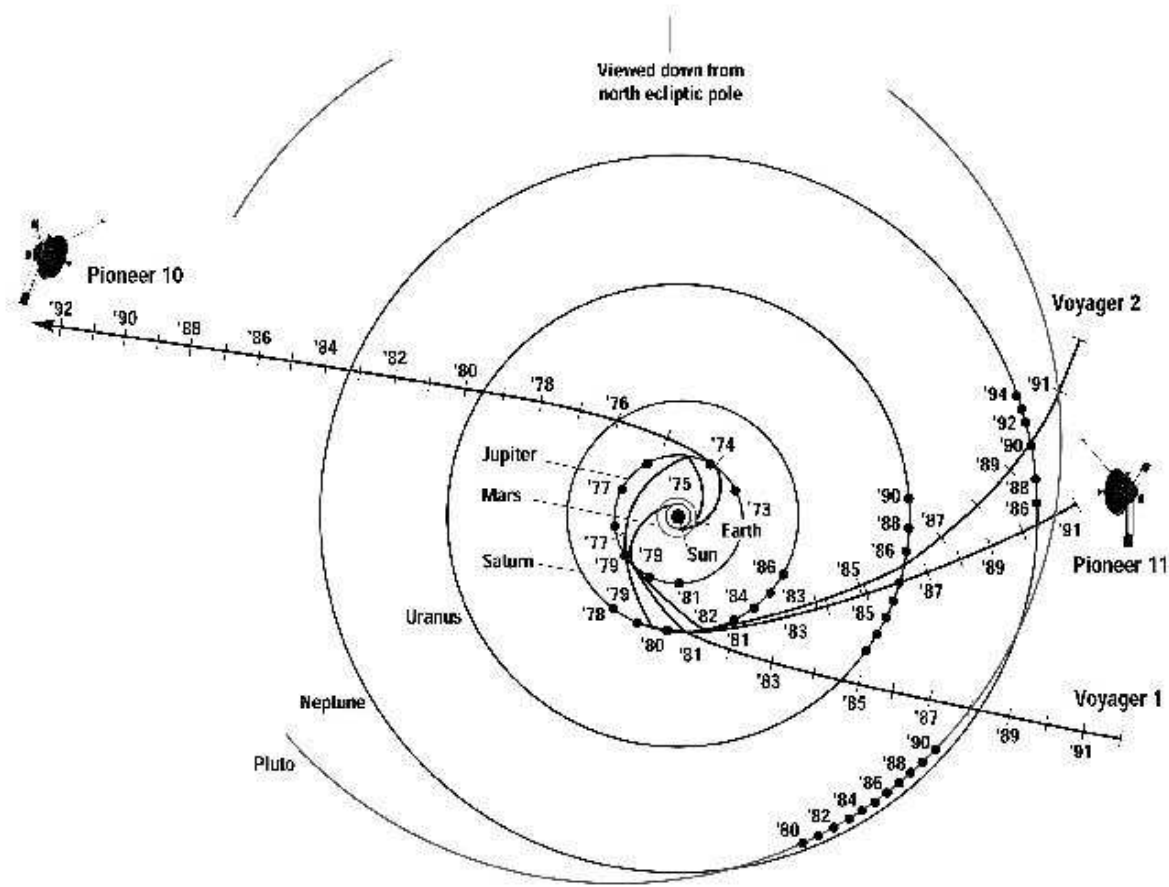,width=160mm}
\end{center}
\caption{Ecliptic pole view of the Pioneer 10, Pioneer 11, and Voyager trajectories. Pioneer 10 is traveling in a direction almost opposite to the galactic center, while Pioneer 11 is heading approximately in the shortest direction to the heliopause. The direction of the solar system's motion in the galaxy is approximately towards the top.  (Digital artwork by T.\ Esposito, NASA ARC Image \# AC97-0036-3.)
\label{fig:pioneer_path}}
\end{figure*}

The Pioneer 10/11 missions, launched on 2 March 1972 (Pioneer 10) and 4
December 1973 (Pioneer 11), were the first to explore the outer solar
system. After Jupiter and (for Pioneer 11) Saturn encounters, the two
spacecraft followed escape hyperbolic orbits near the plane of the ecliptic to opposite sides of the
solar system. (The hyperbolic escape velocities are $\sim$12.25\,km/s
for Pioneer 10 and $\sim$11.6\,km/s for Pioneer 11.) Pioneer 10
eventually became the first man-made object to leave the solar system. See
Fig.~\ref{fig:pioneer_path} for a perspective of the orbits of the
spacecraft and Tables~\ref{I} and \ref{II} for more details on their
missions. The orbital parameters for the crafts are given in
Table~\ref{poeas00479}.

\begin{figure*}[ht!]
\begin{center}
\psfig{figure=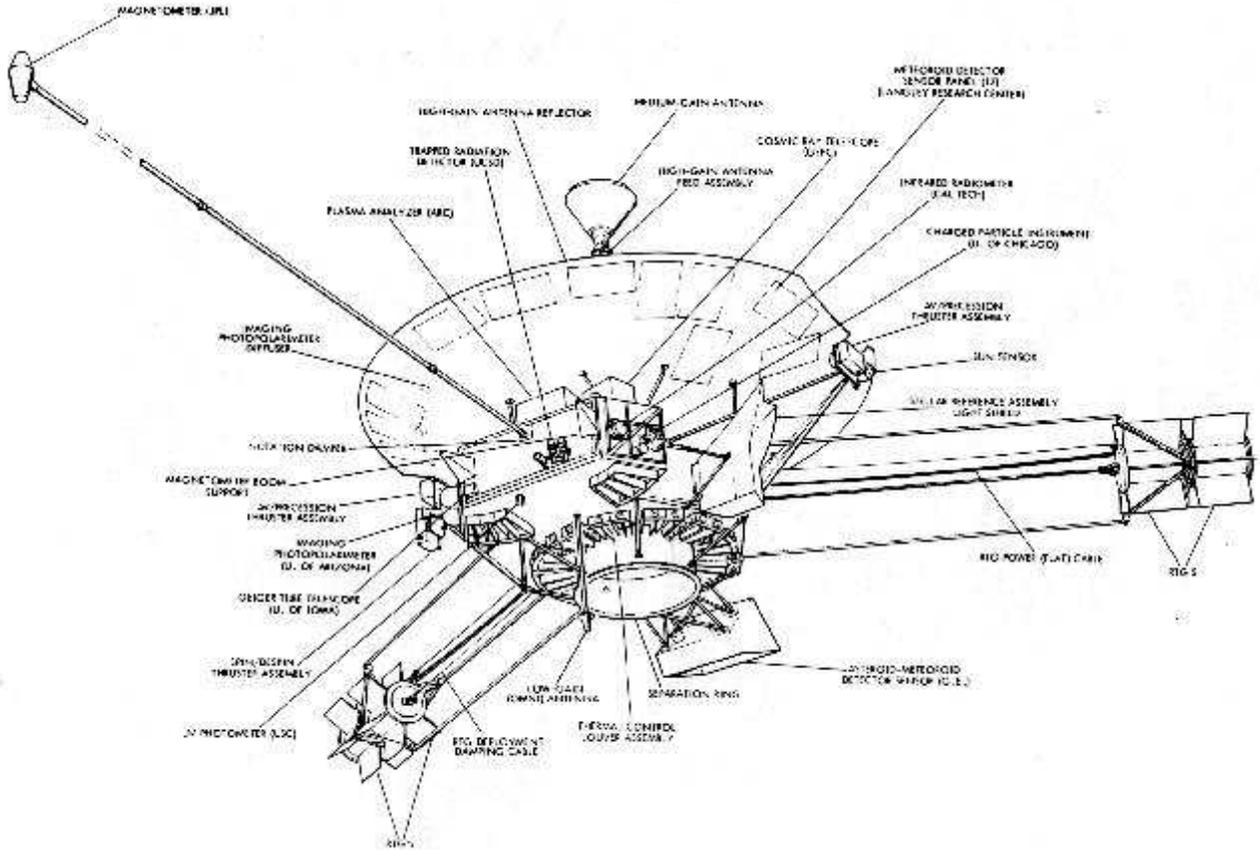,width=170mm}
\end{center}\vskip -10pt
\caption{Drawing of the Pioneer spacecraft. 
\label{fig:thrusters}}
\end{figure*}

The Pioneers were excellent craft with which to perform precise celestial
mechanics experiments due to a combination of many factors,
including their attitude control (spin-stabilized, with a minimum number of attitude correction maneuvers using thrusters), power design [the
Plutonium-238 powered heat-source radioisotope thermoelectric
generators (RTGs) are on extended booms to aid the stability of the craft
and also reduce the effects due to heating], and precise Doppler tracking
(with the accuracy of post-fit Doppler residuals at the level of mHz).
The result was the most precise navigation in deep space to date. See
Fig.~\ref{fig:thrusters} for a design drawing of the spacecraft.

\begin{table*}[ht!]
\begin{center}
\caption{Pioneer 10/11 mission status.
\label{I}}
\begin{tabular}{|c|c|c|} 
\multicolumn{2}{c}{}\\ \hline
 & Pioneer 10 & Pioneer 11 \\ \hline
\hline
Launch& 2 March 1972 & 5 April 1973 \\\hline
Planetary encounters &~Jupiter: 4 Dec 1973~ & ~Jupiter: 2 Dec 1974~ \\ 
 & & Saturn: 1 Sep 1979\\\hline
Last data received & 27 April 2002 & 1 October 1990\\ 
at heliocentric distance & $\approx 80.2$\,AU & $\approx 30$\,AU\\\hline
Direction of motion & Star Aldebaran & Constell.\ of Aquila\\
~time to reach destination ~ & $\approx 2$ million years & $\approx4$
million years\\\hline
\multicolumn{3}{c}{}\\[-10pt] 
\end{tabular}
\end{center}
\end{table*}

\begin{table}[h!]
\begin{center}
\caption{Position of Pioneer 10 on 1 January 2005 (data taken from the
Jet Propulsion Laboratory's Horizons system {\tt
{<}http://ssd.jpl.nasa.gov/horizons.html{>}}).
\label{II}}
\begin{tabular}{|c|c|} 
\multicolumn{2}{l}{}\\ \hline 
Distance from Sun & 87.06\,AU \\\hline
Position (lat., long.) & (3.0$^\circ$,
78.0$^\circ$)\\\hline 
~Speed relative to the Sun~ & 12.24\,km/s \\\hline
Distance from Earth & $13.14\times 10^9$\,km \\\hline
Round-Trip Light Time & ~$\approx$ 24 hr 21 min~ \\\hline
\end{tabular}
\end{center}
\end{table}

\begin{table}[h!]
\begin{center}
\caption{\label{poeas00479}Orbital parameters for Pioneers 10 and 11 at
epoch 1 January 1987, 01:00:00\,UTC. 
The semi-major axis is $a$, 
$e$ is the eccentricity, 
$I$ is the inclination, 
$\Omega$ is the longitude of the ascending node, 
$\omega$ is the argument of the perihelion, 
$M_0$ is the mean anomaly, 
$f_0$ is the true anomaly at epoch, 
and $r_0$ is the heliocentric radius at the epoch. 
The direction cosines of the spacecraft
position for the axes used are 
$(\alpha, \,\beta, \,\gamma)$. 
These direction cosines and angles 
are with respect to the mean equator and equinox of J2000. The ecliptic
longitude $\ell_0$ and latitude $b_0$ are also listed for an obliquity of
23$^\circ$26$^{'}$21.$^{''}$4119. The numbers in parentheses denote
realistic standard errors in the last digits.}
\vskip 2pt
\begin{tabular}{|c|r|r|}\hline\hline
Parameter & Pioneer 10 & Pioneer 11 \\ \hline
$a$ [km] & $-1033394633(4)$ & $-1218489295(133)$ \\[1pt]
$e$ & $1.733593601(88)$ & $2.147933251(282)$ \\[1pt]
$I$ [deg] & 26.2488696(24) & 9.4685573(140) \\[1pt]
$\Omega$ [deg] & $-3.3757430(256)$ & 35.5703012(799) \\[1pt]
$\omega$ [deg] & $-38.1163776(231)$ & $-221.2840619(773)$ \\[1pt]
$M_0$ [deg] & $259.2519477(12)$ & $109.8717438(231)$ \\[1pt]
$f_0$ [deg] & $112.1548376(3)$ & 81.5877236(50) \\[1pt]
$r_0$ [km] & $5985144906(22)$ & 3350363070(598) \\ [1pt]
$\alpha$ & $0.3252905546(4)$ & $-0.2491819783(41)$ \\[1pt]
$\beta$ & $0.8446147582(66)$ & $-0.9625930916(22)$ \\[1pt]
$\gamma$ & $0.4252199023(133)$ & $-0.1064090300(223)$ \\[1pt]
$\ell_0$ [deg] & $70.98784378(2)$ & $-105.06917250(31)$ \\[1pt]
$b_0$ [deg] & $3.10485024(85)$ & $16.57492890(127)$ \\[1pt]
\hline 
\end{tabular} 
\end{center} 
\end{table}

By 1980 Pioneer 10 had passed a distance of $\sim$20 astronomical units
(AU) from the Sun and the acceleration contribution from solar-radiation
pressure on the craft (directed away from the Sun) had decreased to $<4
\times 10^{- 8}$\,cm/s$^2$. At that time an anomaly in the Doppler signal
became evident and persisted.\cite{pioprl} Subsequent analysis of the
radio-metric tracking data from the Pioneer 10/11 spacecraft at distances
between $\sim$20 and 70\,AU from the Sun has consistently indicated the
presence of an anomalous, small, and constant Doppler frequency drift of
$\dot{\nu}=(5.99 \pm 0.01) \times 10^{-9}$\,Hz/s. The drift can be
interpreted as being due to a constant acceleration of the spacecraft of
$a_P = (8.74 \pm 1.33)\times 10^{-8}$\,cm/s$^2$ directed toward the
Sun.\cite{pioprd} The nature of this anomaly remains unknown, and no
satisfactory explanation of the anomalous signal has been found. This signal has become known as the Pioneer anomaly.\cite{foot1}

Although the most obvious explanation would be that there is a systematic
origin to the effect, perhaps generated by the spacecraft themselves from
excessive heat or propulsion gas leaks,\cite{murphy,katz,scheffer} none has been found.\cite{usmurphy,uskatz,mpla} Our inability to explain the
anomalous acceleration of the Pioneer spacecraft has contributed to a
growing discussion about its origin, especially because an independent
verification of the anomaly's existence has been performed.\cite{markwardt}

Attempts to verify the anomaly using other spacecraft have proven
disappointing,\cite{pioprd,mpla,ijmpd_02,Nieto_Turyshev_cqg_2004} because
the Voyager, Galileo, Ulysses, and Cassini spacecraft navigation data all
have their own individual difficulties for use as an independent test of
the anomaly. In addition, many of the deep space missions that are currently being planned either will not provide the needed navigational accuracy and trajectory stability of under $10^{-8}$\,cm/s$^2$ (that is, recent proposals for the Pluto Express and Interstellar Probe missions), or else they will have significant on-board effects that mask the anomaly (the Jupiter Icy Moons Orbiter mission). 
The acceleration regime in which the anomaly was observed diminishes the
value of using modern disturbance compensation systems for a test. For
example, the systems that are currently being developed for the LISA (Laser Interferometric Space Antenna) and LISA Pathfinder missions are designed to operate in the presence of a very low frequency acceleration noise (at the millihertz level), while the Pioneer anomalous acceleration is a strong constant bias in the Doppler frequency data. In addition, currently available accelerometers are a few orders of magnitude less sensitive than is needed for a test. \cite{Nieto_Turyshev_cqg_2004,ijmpd_02,cospar04,mexico04,stanford04} Should the anomaly be a fictitious force that universally affects frequency standards,\cite{pioprd} the use of accelerometers will shed no light on what is the true nature of the observed anomaly. 

A comprehensive test of the anomaly requires an escape hyperbolic
trajectory,\cite{pioprd,mpla,Nieto_Turyshev_cqg_2004} which makes a number
of advanced missions (such as LISA, the Space Test of Equivalence Principle, and LISA Pathfinder) less able to test properties of the detected anomalous acceleration. Although these missions all have excellent scientific goals and technologies, they will be in a less advantageous position to conduct a precise test of the detected anomaly because of their orbits. These analyses of the capabilities of other spacecraft currently in operation or in design lead to the conclusion that none could perform an independent verification of the anomaly and that a dedicated test is needed.\cite{Nieto_Turyshev_cqg_2004,ijmpd_02,cospar04,mexico04} 
Because the Pioneer anomaly has remained unexplained, interest in it has
grown.\cite{cospar04,mexico04,stanford04} Many novel ideas have been
proposed to explain the anomaly, such as a modification of general
relativity, \cite{serge,sanders,moffat} a change in the properties of
light,\cite{light} or a drag from dark matter.\cite{foot} So far, none has
been able to unambiguously explain the anomaly. 

The aim of the following problems is to excite and engage students and to
show them by example how research really works. As a byproduct we hope to
demonstrate that they can take part and understand.

In Sec.~\ref{sec:calc} we begin with a few calculations that are
designed to provide intuition into the dynamics of the problem. Then we
concentrate on the error analysis that demonstrated the existence of the
anomaly in the data. In Sec.~\ref{ext-systema} we will consider possible
sources of errors in the measurement of the Pioneer anomalous acceleration
that have origins external to the spacecraft. Section~\ref{int-systema} will
consider likely sources of errors that originate on-board of the
spacecraft. We will discuss the anticipated effect of computational errors
in Sec.~\ref{Int_accuracy} and summarize the various contributions in
Sec.~\ref{budget}. In Appendix~\ref{allan_dev} we will discuss the Allan
deviation -- a quantity useful in evaluating the stability of the reference frequency standards used for spacecraft navigation. 

\section{Effect of the anomaly on the Pioneers' trajectories}
\label{sec:calc}

For these problems we can use the approximation of ignoring the angular
momentum in the hyperbolic orbits, and treat the velocities as radial. 

{\bf Problem 2.1}. Given the values of the orbital parameters of Pioneer 10
on 1 January 2005 (distance $r_0$ and velocity $v_0$) in Table~\ref{II},
what would be the final velocity of Pioneer 10 assuming that there is no
anomaly? 

{\bf Solution 2.1}. The terminal escape velocity can be calculated from 
conservation of energy:
\begin{equation} v_\infty = \sqrt{v_0^2 - 2GM_\odot/r_0} =
11.38\,\mathrm{km/s}. 
\end{equation}

{\bf Problem 2.2}. Assume that the anomaly continues and is
a constant. At what distance from the Sun would the acceleration of the
anomaly be equal to that of gravity? 

{\bf Solution 2.2}.
We equate the anomalous and Newtonian gravitational forces and
find
\begin{equation}
d = \sqrt{GM_\odot/a_P} \approx 2,600\,\mathrm{AU}.
\end{equation} 

{\bf Problem 2.3}. Assume the anomaly is due to a physical force. The time interval over which the Pioneer 10 data was analyzed was 11.5\,years (1987 to 1998.5). Because of the anomaly, what is the approximate shortfall in distance which Pioneer 10 traveled during this time? 

{\bf Solution 2.3}. Simple mechanics yields
\begin{equation}
\Delta d = \frac{1}{2} a_P t^2 \approx 57,500\,{\mathrm{km}}.
\label{at2}
\end{equation}

{\bf Problem 2.4}. If the anomaly were to continue as a constant out to deep space and be due to a force, how far would Pioneer 10 travel before it reaches zero velocity and starts to fall back in to the Sun? When will the Pioneer reach zero velocity?

{\bf Solution 2.4}. As seen in Problem 2.1, the gravitational
potential energy in deep space due to the Sun is small compared to the
kinetic energy of the Pioneer. Therefore, we can ignore it compared 
to the kinetic energy needed to be turned into potential energy.
(This approximation slightly overestimates the time and distance.) If we
use the data from Table~\ref{II}, we can find the time by solving 
\begin{equation}
v_f = 0 = v_0 - a_P t.
\end{equation}
The solution is $t = 1.40 \times 10^{13}\,{\rm s} =445,000$\,yr. 
The distance traveled would be given by 
\begin{equation}
D = r_0 + v_0t - \frac{1}{2}a_Pt^2 \approx 573,300\,\mathrm{AU} =
9\,\mbox{light years.} 
\label{ap2}
\end{equation}
This distance would be well on the Pioneer 10's way to Aldebaran (see
Table~\ref{I}). Because this distance is very large compared to
$r_0=87.06$\,AU, this result verifies the validity of the
approximation of ignoring the gravitational potential energy. 

{\bf Problem 2.5}. What if the Pioneer anomaly is caused by heat from
the RTGs? As stated, they are powered by Plutonium 238, which has a half
life of 87.74 years. (See Sec.~\ref{ext-systema} for more details on the
RTGs.) What would be the change in the final escape velocity of the Pioneers
from this cause?

{\bf Solution 2.5}.
The acceleration would be decreasing exponentially due to the radioactive decay of the $^{238}$Pu.
Therefore, the change in the velocity would be 
\begin{eqnarray}
\Delta v &=& a_P \int^\infty_0 dt~
\exp[ -t(\ln \, 2)/(87.74 ~\mathrm{yr})] \nonumber\\
&=& a_P (87.74 ~ \mathrm{yr})/\ln \,2 = 3.5 \,\mathrm{m/s}.
\end{eqnarray}

\section{\label{ext-systema} 
Sources of acceleration noise external to the craft}

External forces can contribute to all three vector components of spacecraft
acceleration. However, for a rotating spacecraft, such as the
Pioneers, the forces generated on board produce an averaged contribution only
along the direction of the axis of rotation. The difference in the second
case is due to the fact that the two non-radial components (those that are
effectively perpendicular to the spacecraft spin) are averaged out by the
rotation of the spacecraft. Furthermore, non-radial spacecraft accelerations
are difficult to observe by the Doppler technique, which measures the
velocity along the Earth-spacecraft line of sight. Although we could in
principle set up complicated models to predict effects due to all or each of
the sources of navigational error, often the uncertainty of the models is
too large to make them useful, despite the significant effort required to
use them. A different approach is to accept our ignorance about these
non-gravitational accelerations and assess to what extent they could
contribute to spacecraft acceleration over the time scales of all or part of
the missions. 

In this section we will discuss possible sources of acceleration noise 
originated externally to the spacecraft that might significantly affect the
navigation of the Pioneers 10 and 11. We first consider the forces that
affect the spacecraft motion, such as those due to solar-radiation pressure
and solar wind pressure. We then discuss the effects on the propagation of
the radio signal that are from the solar corona, electromagnetic 
forces, and the phase stability of the reference atomic clocks. 

{\bf Problem 3.1}. There is an exchange of momentum when solar photons
impact the spacecraft and are either absorbed or reflected. The models of
this process take into account various parts of the spacecraft surface
exposed to solar radiation, primarily the high-gain antenna, predict an
acceleration directed away from the Sun as a function of the orientation of
the spacecraft and its distance to the Sun. The effect of the solar
radiation pressure depends on the optical properties of the spacecraft
surface (that is, the absorptivities, reflectivities and emissivities of
various materials used to build the spacecraft), and the effective areas of
the spacecraft part. The effect can be distinguished from the $1/r^2$ law
due to gravity because the direction between the Sun and the effective
spacecraft surface varies. 

For navigational purposes, we determine the magnitude of the solar-pressure
acceleration at various orientations using Doppler data. The following
equation provides a good model for analyzing the effect of solar radiation
pressure on the motion of distant spacecraft. It is included by most of the
programs used for orbit determination:
\begin{equation}
a_{\rm sp}(r)=\frac{\kappa f_\odot}{m\,c}
\frac{ A\cos\theta(r)}{r^2},
\label{eq:srp}
\end{equation}
where $f_\odot=1367\,{\rm W/m}^{2}$(AU)$^2$ is the effective Stefan-Boltzmann
temperature or solar radiation constant at 1\,AU from the Sun, $A$ is the
effective area of the craft as seen by the Sun, and $c$ is the speed of
light. The angle $\theta$ is the angle between the axis of the antenna and
the direction of the Sun, $m$ is the mass of the spacecraft, and $r$ is the
distance from the Sun to the spacecraft in AU. 

For the Pioneers the effective area of the spacecraft, $A$, can be taken to
be the antenna dish of radius 1.37\,m. (See Fig.~\ref{fig:thrusters} for
more information.) The quantity $m$ is the mass of the Pioneers when half
of the hydrazine thruster fuel is gone (241\,kg). Finally, $\kappa$ in
Eq.~(\ref{eq:srp}) is the effective absorption/reflection coefficient of the
spacecraft surface, which for Pioneer 10 was measured to be $\kappa=1.71$. A
similar value was obtained for Pioneer 11. 

Estimate the systematic error from solar-radiation pressure on the Pioneer
10 spacecraft over the interval from 40 to 70.5\,AU, and for Pioneer 11 from
22.4 to 31.7\,AU. 

{\bf Solution 3.1}. With the help of Eq.~(\ref{eq:srp}) we can estimate
that when the craft reached 10\,AU, the solar radiation acceleration was
$18.9\times 10^{-8}$\,cm/s$^2$ decreasing to $0.39 \times 10^{-8}$\,cm/s$^2$
for 70\,AU. Because this contribution falls off as $r^{-2}$, it can bias the
Doppler determination of a constant acceleration. By taking the average of
the $r^{-2}$ acceleration curves over the Pioneer distance, we can estimate
the error in the acceleration of the spacecraft due to solar-radiation
pressure. This error, in units of 10$^{-8}$\,cm/s$^2$, is 0.001 for Pioneer
10 over the interval from 40 to 70.5\,AU, and 0.006 for Pioneer 11 over the
interval from 22.4 to 31.7\,AU. 

{\bf Problem 3.2}. Estimate the effect of the solar wind on the
Pioneer spacecraft. How significant is it to the accuracy of the Pioneers'
navigation? 

{\bf Solution 3.2}. The acceleration caused by the solar wind has the same
form as Eq.~(\ref{eq:srp}), with $f_\odot/c$ replaced by $m_pv^2n$, where
$m_p$ is the mass of proton, $n \approx5$\,cm$^{-3}$ is the proton density
at 1\,AU, and $v\approx400$\,km/s is the speed of the wind. Thus, 
\begin{eqnarray} 
\sigma_{\rm sw}(r) &=& 
\frac{\kappa\,m_pv^2\,n}{m} 
\frac{A\cos\theta(r)}{r^2} \nonumber\\
&\approx& 4.4\times10^{-11}
\Big(\frac{20\,\rm AU}{r}\Big)^2\,{\rm cm/s}^2.
\label{eq:sw}
\end{eqnarray}
We will use the notation $\sigma$ to denote accelerations
that contribute only to the error in determining the magnitude of the
anomaly.

Because the density of the solar wind can change by as much as 100\%,
the exact acceleration is unpredictable. Even if we make the conservative
assumption that the solar wind contributes only 100 times less force than
solar radiation, its smaller contribution is completely negligible. 

{\bf Problem 3.3}. Radio observations in the solar system are affected by
the electron density in the outer solar corona. Both the electron density
and density gradient in the solar atmosphere influence the propagation of
radio waves through this medium. Thus, the Pioneers' Doppler
radio-observations were affected by the electron density in the
interplanetary medium and outer solar corona. In particular, the one way
time delay associated with a plane wave passing through the solar corona is
obtained by integrating the group velocity along the ray's path, $\ell$:
\begin{eqnarray}
\Delta t &=& - \frac{1}{2c\,n_{\rm crit}(\nu)}
\int_\oplus^{SC}d\ell n_e(t, \rv), \label{eq:sol_plasma1}\\
n_{\rm crit}(\nu) &=& 1.240 \times 10^4 
\Big(\frac{\nu}{1\,{\rm MHz}}\Big)^2\,{\rm cm}^{-3},
\label{eq:sol_plasma}
\end{eqnarray}
where $n_e(t, \rv)$ is the free electron density in the solar plasma, and
$n_{\rm crit}(\nu)$ is the critical plasma density for the radio carrier
frequency $\nu$. 

We see that to account for the plasma contribution, we should know the
electron density along the path. We usually write the electron density,
$n_e$, as a static, steady-state part, $\overline{n}_e(\bf{r})$, plus a
fluctuation $\delta n_e(t, \rv)$: 
\begin{equation}
n_e(t, \rv)= \overline{n}_e(\rv)+ \delta n_e(t, \rv). 
\end{equation}
The physical properties of the second term are difficult to quantify. But its effect on Doppler observables and,
therefore, on our results is small. On the contrary, the steady-state corona
behavior is reasonably well known and several plasma models can be found in
the literature.\cite{pioprd,corona1,corona2,corona3,corona4} 

To study the effect of a systematic error from propagation of the Pioneer
carrier wave through the solar plasma, we adopted the following model for the
electron density profile,
\begin{eqnarray}
n_e(t, \rv)&=& 
A\Big(\frac{R_\odot}{r}\Big)^2 +\nonumber\\
&& +~
B\Big(\frac{R_\odot}{r}\Big)^{2.7}
e^{-[\frac{\phi}{\phi_0}]^2}+
C\Big(\frac{R_\odot}{r}\Big)^6,
\label{corona_model_content1}
\end{eqnarray}
where $r$ is the heliocentric distance to the immediate ray trajectory and
$\phi$ is the helio-latitude normalized by the reference latitude of
$\phi_0=10^\circ$. The parameters $r$ and $\phi$ are determined from the
signal propagation trajectory. The parameters $A$, $B$, $C$ describe the
solar electron density. (They are commonly given in cm$^{-3}$.) 

Define $\Delta d=c\Delta t$ to be the delay of radio signals due to the
solar corona contribution and use
Eqs.~\eqref{eq:sol_plasma1}--\eqref{corona_model_content1} to derive an
analytical expression for this quantity. 

{\bf Solution 3.3}.
The substitution of Eq.~(\ref{corona_model_content1}) into
Eq.~(\ref{eq:sol_plasma1}) results in the following steady-state solar
corona contribution to the earth-spacecraft range model that was used in the Pioneer analysis:\cite{pioprd}
\begin{eqnarray} 
\Delta d &=& - \Big(\frac{\nu_0}{\nu}\Big)^2\eta \bigg[
A\Big(\frac{R_\odot}{\rho}\Big)F +\nonumber \\
&& +~ B\Big(\frac{R_\odot}{\rho}\Big)^{1.7}
e^{-[\frac{\phi}{\phi_0}]^2}+
C\Big(\frac{R_\odot}{\rho}\Big)^{5}\bigg].
\label{corona_model}
\end{eqnarray}
Here $\eta=R_\odot/2\,n_{\tt crit}(\nu_0)$ is the units conversion factor from cm$^{-3}$ to meters, $\nu_0$ is the reference frequency ($\nu_0=2295$\,MHz for Pioneer 10), $\nu$ is the actual frequency of the radio wave, and $\rho$ is the impact parameter with respect to the Sun. 
The function $F$ in Eq.~(\ref{corona_model}) is a light-time correction factor that was obtained during integration of Eqs.~(\ref{eq:sol_plasma1}) and (\ref{corona_model_content1}) as:\cite{pioprd} 
\begin{eqnarray}
F=F(\rho, r_T, r_E) &=& \frac{1}{\pi}\Bigg\{ 
\arctan\Big[\frac{\sqrt{r_T^2-\rho^2}}{\rho}\Big] +\nonumber\\
&& +~
\arctan \Big[\frac{\sqrt{r_E^2-\rho^2}}{\rho}\Big]\Bigg\},
\label{eq:weight_doppler*}
\end{eqnarray}
where $r_T$ and $r_E$ are the heliocentric radial distances to
the target spacecraft and to the Earth, respectively. 

{\bf Problem 3.4}. The analyses of the Pioneer anomaly\cite{pioprd} took
the values for $A$, $B$, and $C$ obtained from the Cassini mission on its
way to Saturn and used them as inputs for orbit determination purposes: $(A,\,B,\,C)\,\eta = (6.0\times 10^3,\, 2.0\times 10^4, \,0.6\times 10^6)$, all in meters.

Given Eq.~(\ref{corona_model}) derived in Problem~3.3 and the
values of the parameters $A$, $B$, and $C$, estimate the acceleration error due to the effect of the solar corona on the propagation of radio waves between the Earth and the spacecraft. 

{\bf Solution 3.4}.
The correction to the Doppler frequency shift is obtained from
Eq.~(\ref{corona_model}) by simple time differentiation. (The impact
parameter depends on time as $\rho=\rho(t)$ and may be expressed in terms of the relative velocity of the spacecraft with respect to the Earth, $v\approx 30$\,km/s). Use Eq.~(\ref{corona_model}) for the steady-state solar corona effect on the radio-wave propagation through the solar plasma. The effect of the solar corona is expected to be small on the Doppler frequency shift -- our main observable. The time-averaged effect of the corona on the propagation of the Pioneers' radio-signals is found to be small, of order 
\begin{equation}
\sigma_{\rm corona} = \pm 0.02 \times 10^{-8}\,\mathrm{cm/s}^2. 
\end{equation}
This small result is expected from the fact that most of the data used for
the Pioneer analysis were taken with large Sun-Earth-spacecraft angles. 

{\bf Problem 3.5}. The possibility that the Pioneer spacecraft can hold a
charge and be deflected in its trajectory by Lorentz forces was a concern
due to the large magnetic field strengths near Jupiter and Saturn (see
Figs.~\ref{fig:thrusters} and \ref{fig:pioneer_path}).\cite{foot2} The
magnetic field strength in the outer solar system is $\le10^{-5}$\,Gauss,
which is about a factor of
$10^5$ times smaller than the magnetic field strengths measured by the
Pioneers at their nearest approaches to Jupiter: 0.185\,Gauss for Pioneer 10
and 1.135\,Gauss for Pioneer 11. Furthermore, data from the Pioneer 10
plasma analyzer can be interpreted as placing an upper bound of $0.1 \mu$C
on the positive charge during the Jupiter encounter.\cite{null}

Estimate the upper limit of the contribution of the electromagnetic force
on the motion of the Pioneer spacecraft in the outer solar system. 

{\bf Solution 3.5}. We use the standard relation, ${\mathbf F} = q
\mathbf{v} \times {\mathbf B}$, and find that the greatest force would be on
Pioneer 11 during its closest approach,
$<20\times 10^{-8}$\,cm/s$^{2}$. However, once the spacecraft reached the
interplanetary medium, this force decreased to 
\begin{equation}
\sigma_{\rm Lorentz} \lesssim 2 \times 10^{-12}\,\mathrm{cm/s}^{2},
\end{equation} 
which is negligible. 

{\bf Problem 3.6}. Long-term frequency stability tests are conducted with
the exciter/transmitter subsystems and the Deep Space Network's
(DSN)\cite{dsn} radio-science open-loop subsystem. An uplink signal
generated by the exciter is converted at the spacecraft antenna to a downlink frequency. The downlink signal is then passed through the RF-IF
down-converter present at the DSN antenna and into the radio science receiver
chain. This technique allows the processes to be synchronized in the DSN
complex based on the frequency standards whose Allan deviations are the
order of $\sigma_y \sim 10^{-14}$--$10^{-15}$ for integration times in the
range of 10\,s to $10^3$\,s. (The Allan deviation is the variance in the
relative frequency, defined as $y=\Delta\nu/\nu$, or
$\sigma_y\sim\sigma_\nu/\nu$. See the discussion in
Appendix~\ref{allan_dev}.) For the S-band frequencies of the Pioneers, the
corresponding variances are $1.3 \times 10^{-12}$ and $1.0 \times 10^{-12}$,
respectively, for a $10^3$\,s Doppler integration time. 

The influence of the clock stability on the detected acceleration, $a_P$,
can be estimated from the reported variances for the clocks, $\sigma_y$. The standard single measurement error on acceleration is derived from the time derivative of the Doppler frequency data as $(c\sigma_y)/\tau$, where the variance, $\sigma_y$, is calculated for 1000\,s Doppler integration time and $\tau$ is the signal averaging time. This relation provides a good rule of thumb when the Doppler power spectral density function obeys a $1/f$ flicker noise law, which is approximately the case when plasma noise dominates the overall error in the Doppler observable.

Assume a worst case scenario, where only one clock was used for the entire
11.5 years of data. Estimate the influence of that one clock on the reported accuracy of the detected anomaly, $a_P$. Combining
$\sigma_y=\Delta\nu/\nu_0$, the fractional Doppler frequency shift for a reference frequency of $\nu_0\approx 2.295$\,GHz, with the estimate for the variance, $\sigma_y =1.3\times 10^{-12},$ yields the upper limit for the frequency uncertainty introduced in a single measurement by the instabilities in the atomic clock:
$\sigma_\nu=\nu_0\sigma_y=2.98\times10^{-3}$\,Hz for a $10^3$ Doppler
integration time.

Obtain an estimate for the total effect of phase and frequency stability of clocks to the Pioneer anomaly value. How significant is this effect?

{\bf Solution 3.6}. To obtain an estimate for the total effect,
recall that the Doppler observation technique is essentially a continuous
count of the total number of complete frequency cycles during
the observational time. The Pioneer measurements were made with duration
$\approx 10^3$\,s. Therefore, within a year we could have as many as
$N\approx 3,150$ independent single measurements of the clock.
The values for $\sigma_y$ and $N$ yield an upper limit for the contribution of
the atomic clock instability on the frequency drift:
$\sigma_{\rm clock} = {\sigma_\nu}/{\sqrt{N}} \approx 5.3\times
10^{-5}$\,Hz/yr. The observed $a_P$ corresponds to a
frequency drift of about 0.2\,Hz/yr, which means that the error in $a_P$
is about 
\begin{equation}
\sigma_{\rm freq} = 0.0003 \times 10^{-8}\,\mathrm{cm/s}^2, 
\end{equation}
which is negligible. 

\section{\label{int-systema} On-board sources of acceleration noise}

In this section forces generated by on-board spacecraft systems that could
contribute to the constant acceleration, $a_P$, are considered. The on-board
mechanisms discussed are the radio beam reaction force, the RTG heat
reflecting off the spacecraft, the differential emissivity of the RTGs,
the expelled Helium produced within the RTG, the thruster gas
leakage, and the difference in experimental results from the two
spacecraft.

{\bf Problem 4.1}.
The Pioneers have a total nominal emitted radio power of 8\,W. The power is parametrized as 
\begin{equation}
P_{\rm rp} =\int_0^{\theta_{\max}} d\theta \,\sin\theta {\cal
P}(\theta),
\end{equation}
where ${\cal P}(\theta)$ is the antenna power distribution.
The radiated power is kept constant in time, independent of the coverage
from ground stations. That is, the radio transmitter is always on, even when signals are not received by a ground station. 

The recoil from this emitted radiation produces an acceleration 
bias,\cite{bias} $b_{\rm rp}$, on the spacecraft away from the Earth of
\begin{equation}
b_{\rm rp}= \frac{\beta \,P_{\rm rp}}{m c}, 
\label{eq:rp}
\end{equation}
where $m$ is taken to be the Pioneer mass when half the fuel is gone,
241\,kg, and $\beta$ is the fractional component of the radiation
momentum that is in the direction opposite to $a_P$:
\begin{equation}
\beta =\frac{1}{P_{\rm rp}}
{\int_0^{\theta_{\rm max}} d\theta \sin\theta \cos\theta 
{\cal P}(\theta)}.
\label{radiopower}
\end{equation}

The properties of the Pioneer downlink antenna patterns are well known. The gain is given as $(33.3 \pm 0.4)$\,dB at zero (peak) degrees. The intensity is down by a factor of 2 ($-3$\,dB) at $1.8^\circ$. It is down a factor of 10 ($-10$\,dB) at $2.7^\circ$ and down by a factor of 100 ($-20$\,dB) at $3.75^\circ$. (The first diffraction minimum is at a little over $4^\circ$.) Therefore, the pattern is a very good conical beam. 

Estimate the effect of the recoil force due to the emitted radio-power on
the craft. What is the uncertainty in this estimation?

{\bf Solution 4.1}.
Because $\cos 3.75^\circ = 0.9978$, we can take $\beta = (0.99 \pm
0.01)$, which yields $b_{\rm rp}=1.10$. To estimate the uncertainty, we take
the error for the nominal 8\,W power to be given by the 0.4\,dB antenna
error (0.10), and obtain
\begin{equation}
a_{\rm rp} =b_{\rm rp} \pm \sigma_{\rm rp}
=(1.10 \pm 0.10)\times 10^{-8}\,{\rm cm/s}^2.
\label{eq:rad-beam}
\end{equation}
\begin{figure}[h]
\begin{center} 
\epsfig{figure=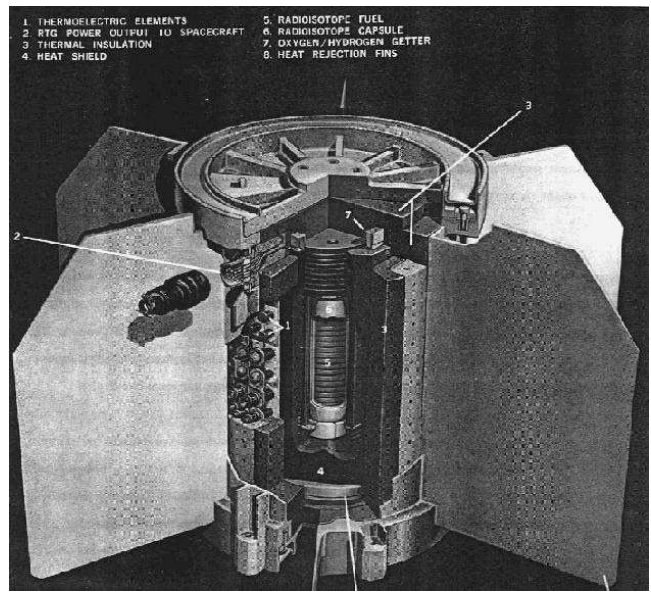,width=85mm}
\end{center}
\caption{A design picture of the SNAP 19 RTGs. Note that for better heat
radiation, this final design had much larger fins than those in the earlier
spacecraft concept of Fig.~\ref{fig:thrusters}. 
\label{fig:snap19rtg}}
\end{figure}

\begin{table*}[ht!]
\begin{center}
\caption{Thermal system and on-board power for Pioneers 10 and 11. The core of the system is four SNAP-19 RTGs (SNAP stands for Space Nuclear Auxiliary Power). These RTGs are fueled with Plutonium 238 that undergoes nuclear $\alpha$-decay via ${}_{94}$Pu$^{238}
\rightarrow{}_{92}$U$^{234}+{}_2$He$^4$, with a half life of 87.74\,yr. The efficiency of the relevant heat-to-electric power conversion is about 5 to 6\%.
\label{IV}}
\begin{tabular}{|l|c|} 

\multicolumn{1}{l}{}\\[0pt] \hline 
\multicolumn{2}{|l|}
{Power available:}\\ 
\multicolumn{2}{|l|}
{ \hskip 20pt $\triangleright$~
before launch electric total 165\,W (by 2001 $\approx 61$\,W)}\\
\multicolumn{2}{|l|}
{ \hskip 20pt $\triangleright$ ~
needs 100\,W to power all systems 
($\in$ 24.3\,W science instruments)}\\\hline 

\multicolumn{2}{|l|}
{ Heat provided:}\\ 
\multicolumn{2}{|l|}
{\hskip 20pt $\triangleright$ ~ before launch thermal fuel total 2580\,W
(by 2001 $\approx 2050$\,W)}\\
\multicolumn{2}{|l|}
{ \hskip 20pt $\triangleright$~ electric heaters; 12 one-W RHUs}\\

\multicolumn{2}{|l|}
{ \hskip 20pt $\triangleright$ ~ heat from the instruments (dissipation of
70 to 120\,W) \large}\\\hline 

\multicolumn{2}{|l|}
{ Excess power/heat: 
~~~~~if electric power was $>$ 100\,W $\Rightarrow$ }\\ 
\multicolumn{2}{|l|}
{ \hskip 20pt $\triangleright$ ~
thermally radiated into space by a shunt-resistor radiator, or}\\
\multicolumn{2}{|l|}
{ \hskip 20pt $\triangleright$ ~
charge a battery in the equipment compartment }\\\hline 
\multicolumn{2}{|l|}
{Thermal control: 
~~~~~~~~to keep temperature within (0--90)\,F}\\ 
\multicolumn{2}{|l|}
{ \hskip 20pt $\triangleright$~
thermo-responsive louvers: completely shut/fully open for 40\,F/85\,F} \\
\multicolumn{2}{|l|}
{ \hskip 20pt $\triangleright$~
insulation: multi-layered aluminized mylar and kapton blankets} \\\hline
\end{tabular}
\end{center}
\end{table*}

{\bf Problem 4.2}. It has been argued that the anomalous
acceleration is due to anisotropic heat reflection off of the back of the
spacecraft high-gain antennas, the heat coming from the RTGs. Before launch, the four RTGs had a total thermal fuel inventory of 2580\,W ($\approx 2050$\,W in 2001). They produced a total electrical power of 165\,W ($\approx 61$\,W in 2001). (In Table~\ref{IV} a description of the overall power system is given.) Thus, by 2002 approximatelly 2000\,W of RTG heat still had to be dissipated. Because only $\approx63$\,W of directed power could have explained the anomaly, in principle, there was enough power to explain the anomaly. 

The main bodies of the RTGs are cylinders and are grouped in two packages
of two. Each package has two cylinders, end to end, extending away from
the antenna. Each RTG has six fins separated by equal angles of
$60^\circ$ that go radially out from the cylinder. The approximate
dimensions for the RTGs are: the cylinder and fin lengths are 28.2 and
26.0\,cm, the radius of the central RTG cylinder is 8.32\,cm, and the RTG
fin width as measured from the surface of the cylinder to the outer fin tip is 17.0\,cm (see Fig.~\ref{fig:snap19rtg}).

The RTGs are located at the end of booms, and rotate about the spacecraft in
a plane that contains the approximate base of the antenna. From the closest
axial center point of the RTGs, the antenna is seen nearly ``edge on'' (the
longitudinal angular width is 24.5$^\circ$). The total solid angle subtended
is $\sim 1$--2\% of $4\pi$ steradians. A more detailed calculation yields a
value of 1.5\%. 

Estimate the contribution of the RTG heat reflecting off the spacecraft to
the Pioneer anomaly. Can you explain the observed anomaly with this
mechanism? Use the spacecraft geometry and the resultant RTG radiation
pattern. 

{\bf Solution 4.2}.
There are two reasons that preclude this mechanism. One is the spacecraft
geometry. Even if we take the higher bound of 2\% of 4$\pi$ steradians
solid angle as the fraction of solid angle covered by the antenna, the
equivalent fraction of the RTG heat could provide at most $\approx 40$\,W.
The second reason is the RTGs' radiation pattern. Our estimate was based on
the assumption that the RTGs are spherical black bodies, which they are not. 

In fact, the fins are ``edge on'' to the antenna (the fins point
perpendicular to the cylinder axes). The largest opening angle of the fins
is seen only by the narrow angle parts of the antenna's outer edges. If we
ignore these edge effects, only $\approx2.5$\% of the surface area of the
RTGs is facing the antenna, which is a factor of 10 less than that obtained
from integrating the directional intensity from a hemisphere: $(\int^{\rm
hemi}d\Omega\cos\theta)/(4\pi)=1/4$. Thus, we have only 4\,W of directed
power. 

The force from 4\,W of directed power suggests a systematic bias of 
$\approx 0.55 \times 10^{-8}$\,cm/s$^2$. If we add an uncertainty of the same
size, we obtain a contribution from heat reflection of 
\begin{equation}
a_{\rm hr}= (-0.55 \pm 0.55) \times
10^{-8}\,\mbox{cm/s}^2. 
\end{equation}

If this mechanism were the cause, ultimately an unambiguous
decrease in the size of $a_P$ should be observed, because the RTGs'
radioactively produced radiant heat is decreasing. As we noted, the heat
produced is now about 80\% of the original magnitude. In fact, we would
expect a decrease of about $0.75\times 10^{-8}$\,cm/s$^2$ in $a_P$ over the
11.5 year Pioneer 10 data interval if this mechanism were the origin of
$a_P$.

{\bf Problem 4.3}. Another suggestion related to the RTGs is the following:
During the early parts of the missions, there might have been a differential
change of the radiant emissivity of the solar-pointing sides of the RTGs
with respect to the deep-space facing sides. Note that, especially closer in
the Sun, the inner sides were subjected to the solar wind. On the other
hand, the outer sides were sweeping through the solar system dust cloud.
These two processes could result in different levels of degradation of the
optical surfaces of the RTGs. In turn this degradation could result in
asymmetric patterns of heat radiation away from the RTGs in the more/aft
directions along the spin axis. Therefore, it can be argued that such an
anisotropy may have caused the anomaly.

The six fins of each RTG, designed to provide the bulk of the heat rejection
capacity, are fabricated of magnesium alloy.\cite{pioprd} The metal is
coated with two to three mils of zirconia in a sodium silicate binder to
provide a high emissivity $(\approx 0.9)$ and low absorptivity $(\approx
0.2)$ (see Fig.~\ref{fig:snap19rtg}).

Estimate the possible difference in the fore-and-aft emissivities of the
RTGs needed to support this mechanism. Can this mechanism explain the
observed anomaly? Discuss the significance of radioactive decay for this
mechanism.

{\bf Solution 4.3}. Depending on how symmetrically fore-and-aft they
radiated, the relative fore-and-aft emissivity of the alloy would have had
to have changed by $\approx10$\% to account for $a_P$. Given our knowledge of
the solar wind and the interplanetary dust, this amount of a radiant change
would be difficult to explain, even if it had the right sign.

To obtain a reasonable estimate of the uncertainty, consider if one side
(fore or aft) of the RTGs has its emissivity changed by only 1\% with
respect to the other side. In a simple cylindrical model of the RTGs, with
2000\,W power (only radial emission is assumed with no loss out of the
sides), the ratio of the power emitted by the two sides would be
$995/1005=0.99$, or a differential emission between the half cylinders of
10\,W. Therefore, the fore/aft asymmetry toward the normal would be
$10\,{\mathrm{W}} \times \frac{1}{\pi}\int_0^\pi d\phi \sin \phi \approx
6.37$\,W. 

A more sophisticated model of the fin structure starts off from ground zero to determine if we still obtain about the order of 
magnitude.\cite{pioprd} Indeed we do, 6.12\,W, which is the value we use in the following. (We refer the reader to Ref.~\onlinecite{pioprd} for this discussion, because the basic physics already has been discussed.)

If we take 6.12\,W as the uncertainty from the differential emissivity of the RTGs, we obtain an acceleration uncertainty of 
\begin{equation}
\sigma_{\rm de} = 0.85 \times 10^{-8}\,{\mathrm{cm/s}}^2.
\end{equation}
Note that $10\,\sigma_{\rm de}$ almost equals our final result for $a_P$ (see Section~\ref{budget}). This correspondence is the origin of our previous statement that $\approx 10$\% differential emissivity (in the correct direction) would be needed to explain $a_P$. 

Finally, consider the significance of radioactive decay for this
mechanism. The formal statistical error of the determination (before
considering experimental systematics) was small\cite{pioprd} (see
Sec.~\ref{Int_accuracy}). An example is a specific one-day
``batch-sequential filtering''\cite{batch-seq} value for $a_P$, averaged over the entire 11.5\,year interval. This value was
$a_P = (7.77 \pm 0.16)\times 10^{-8}$\,cm/s$^2$, where a batch-sequential
error tends to be large. From radioactive decay, the value of $a_P$ should
have decreased by $0.75\times 10^{-8}$\,cm/s$^2$ over 11.5 years, which is
five times the large batch sequential variance. Even more stringently, this bound is good for all radioactive heat sources. So, what if we were to argue that emissivity changes occurring before 1987 were the cause of the Pioneer effect? There still should have been a decrease in $a_P$ with time since then, which has not been observed. 

{\bf Problem 4.4}. Another possible on-board systematic error is from the
expulsion of the He being created in the RTGs from the $\alpha$-decay of
$^{238}$Pu. 

The Pioneer RTGs were designed so that the He pressure is not totally
contained within the Pioneer heat source over the life of the RTGs. Instead,
the Pioneer heat source contains a pressure relief device that allows the
generated He to vent out of the heat source and into the thermoelectric
converter. The thermoelectric converter housing-to-power output receptacle
interface is sealed with a viton O-ring. The O-ring allows the helium gas
within the converter to be released by permeation into the space environment
throughout the mission life of the Pioneer RTGs. 

The actual fuel (see Fig. \ref{fig:snap19rtg}) is formed into a stack of
disks within a container, each disk having a height of 0.212$''$ and a
diameter of 2.145$''$. With 18 disks in each RTG and four RTGs per mission,
the total volume of fuel is about 904\,cm$^3$. The fuel is a plutonium moly
cermet and the amount of $^{238}$Pu in this fuel is about 5.8\,kg.
With a half life of 87.74\,yr, the rate of He production (from plutonium
$\alpha$-decay) is about 0.77\,g/yr, assuming it all leaves the cermet and
goes into the RTG chamber. To make this mechanism work, He leakage from the
RTGs must be preferentially directed away from the Sun with a velocity large
enough to cause the acceleration. 

Can this mechanism explain the Pioneer anomaly? What is the largest effect
possible with this mechanism? What is the uncertainty? 

{\bf Solution 4.4}.
The operational temperature on the RTG surface at 433\,K implies a helium
velocity of 1.22\,km/s. We use this value in the rocket equation, 
\begin{equation} 
a(t) = -v(t) \frac{d}{dt} [\ln M(t)].
\end{equation} 
We assume that the Pioneer mass corresponds to half of the fuel and
that the gas is all unidirected and find a maximum bound on the possible
acceleration of $1.16 \times 10^{-8}$\,cm/s$^2$. Thus, we can rule out
helium permeating through the O-rings as the cause of $a_P$. 

If we assume a single elastic reflection, we can estimate the 
acceleration bias\cite{bias} in the direction away from the Sun. This
estimate is $(3/4) \sin 30^\circ$ times the average of the heat momentum
component parallel to the shortest distance to the RTG fin. By using this
geometric information, we find the bias would be
$0.31 \times 10^{-8}$\,cm/s$^2$. This bias effectively increases the value
of the solution for $a_P$, which is too optimistic given all the
complications of the real system. Therefore, we can take the systematic
expulsion to be 
\begin{equation}
a_{\rm He} = (0.15 \pm 0.16)\times
10^{-8}\,\mathrm{cm/s}^2. 
\end{equation} 

{\bf Problem 4.5}. The effect of propulsive mass expulsion due to gas
leakage has to be assessed. Although this effect is largely unpredictable,
many spacecraft have experienced gas leaks producing accelerations on the
order of $10^{-7}$\,cm/s$^2$. Gas leaks generally behave differently after
each maneuver. The leakage often decreases with time and becomes
negligible. 

Gas leaks can originate from Pioneer's propulsion system, which is used
for mid-course trajectory maneuvers, for spinning up or down the
spacecraft, and for orientation of the spinning spacecraft. The
Pioneers are equipped with three pairs of hydrazine thrusters, which are
mounted on the circumference of the Earth-pointing high gain antenna. 
Each pair of thrusters forms a thruster cluster assembly with two
nozzles aligned opposite to each other. For attitude control, two
pairs of thrusters can be fired forward or aft and are used to precess the
spinning antenna. The other pair of thrusters is aligned parallel to the rim
of the antenna with nozzles oriented in co- and contra-rotation directions
for spin/despin maneuvers. 

During both observation intervals for the two Pioneers, there were no
trajectory or spin/despin maneuvers. Thus, in this analysis we are 
mainly concerned with precession (that is, orientation or attitude control)
maneuvers only. Because the valve seals in the thrusters can never be
perfect, we ask if the leakages through the hydrazine thrusters could
be the cause of the anomalous acceleration, $a_P$. 

Consider the possible action of gas leaks originating from the
spin/despin thruster cluster assembly. Each nozzle from this pair of
thrusters is subject to a certain amount of gas leakage. But only a
differential leakage from the two nozzles would produce an observable
effect, causing the spacecraft to either spin down or up. 

To obtain a gas leak uncertainty (and we emphasize uncertainty rather
than ``error'' because there is no other evidence), we ask, ``How large a
differential force is needed to cause the spin down or spin up effects that
are observed?'' The answer depends on the moment of inertia about the spin
axis, $I_{\rm z} \approx 588.3$\,kg m${^2}$, the antenna radius,
$R=1.37$\,m, and the observed spin down rates over three time intervals. In
Ref.~\onlinecite{pioprd} the final effect was determined at the end by
considering the spin-downs over the three time periods to be manifestations
of errors that are uncorrelated and are normally distributed around zero
mean. For simplicity, we include here the effect of this analysis in terms
of an overall effective spin-down rate of $\ddot{\theta} =-0.0067$\,rpm/yr. 

Estimate the uncertainty in the value for the Pioneer anomaly from the
possibility of undetected gas leaks.

{\bf Solution 4.5}. If we use the information in Problem 4.5 and take the
antenna radius,
$R$, as the lever arm, we can calculate the differential 
force needed to torque the spin-rate change: 
\begin{equation}
F_{\ddot{\theta}} = 
\frac{I_{\rm z}{\ddot{\theta}}}{R} =0.95 \times
10^{-3}\,{\rm dyn}. 
\label{FthetaI} 
\end{equation}

It is possible that a similar mechanism of undetected gas leakage could
be responsible for the net differential force acting in the direction
along the line of sight. In other words, what if there were some
undetected gas leakage from the thrusters oriented along the spin axis
of the spacecraft that is causing $a_P$? How large would this leakage have
to be? A force of 
\begin{equation}
F_{a_P}= m a_P
=21.11\times10^{-3}\,{\rm dyn}
\end{equation}
($m= 241$\,kg) would be needed to produce our final unbiased value of
$a_P$. Given the small amount of information, we conservatively take as our
gas leak uncertainties the acceleration values that would be produced by
differential forces equal to 
\begin{equation}
F_{\rm gl}\approx \pm \sqrt{2}F_{\ddot{\theta}} =
\pm 1.35 \times 10^{-3}\,{\rm dyn}.
\label{eq:diff}
\end{equation} The argument for the result obtained in Eq.~(\ref{eq:diff})
is that we are accounting for the differential leakages from the two pairs
of thrusters with their nozzles oriented along the line of sight direction.
Equation~(\ref{eq:diff}) directly translates into the acceleration
errors introduced by the leakage during the interval of Pioneer 10 data,
\begin{equation}
\sigma_{\rm gl}=\pm F_{\rm gl}/m = \pm 0.56\times10^{-8}\,{\rm cm/s}^2.
\end{equation}

At this point, we must conclude that the gas leak mechanism for explaining
the anomalous acceleration is very unlikely, because it is difficult to
understand why it would affect Pioneers 10 and 11 by the same amount. We also expect a gas leak would contribute to the anomalous acceleration,
$a_P$, as a stochastic variable obeying a Poisson distribution. Instead,
analyses of different data sets indicate that $a_P$ behaves as a constant
bias rather than as a random variable.\cite{pioprd} 

{\bf Problem 4.6}.
There are two experimental results for the Pioneer anomaly from the two
spacecraft: $7.84 \times 10^{-8}$\,cm/s$^2$ (Pioneer 10) and $8.55 \times
10^{-8}$\,cm/s$^2$ (Pioneer 11). The first result represents the
entire 11.5\,year data period for Pioneer 10; Pioneer 11's
result represents a 3.75\,year data period.

The difference between the two craft could be due to different gas
leakage. But it also could be due to heat emitted from the RTGs. In
particular, the two sets of RTGs have had different histories and so
might have different emissivities. Pioneer 11 spent more time in the
inner solar system (absorbing radiation). Pioneer 10 has swept out
more dust in deep space. Further, Pioneer 11 experienced about twice
as much Jupiter/Saturn radiation as Pioneer 10. However, if the Pioneer effect is real, and not due to an extraneous systematic like heat, these numbers should be approximately equal. 

Estimate the value for the Pioneer anomaly based on these two independent
determinations. What is the uncertainty in this estimation? 

{\bf Solution 4.6}. We can calculate the time-weighted average of the
experimental results from the two craft: $[(11.5)(7.84) +
(3.75)(8.55)]/(15.25) = 8.01$ in units of $10^{-8}$\,cm/s$^2$. This result
implies a bias of
$b_{\rm 2,\,craft}=0.17\times10^{-8}$\,cm/s$^2$ with respect to the Pioneer
10 experimental result $a_{P({\rm exp})}$ (see Eq.~(\ref{apexp}) below). In addition, we can take this
number to be a measure of the uncertainty from the separate spacecraft
measurements, so the overall quantitative measure is 
\begin{eqnarray}
a_{\rm 2,\,craft}&=& b_{\rm 2,\,craft}\pm \sigma_{\rm 2,\,craft}  \nonumber\\
&=& (0.17 \pm 0.17)
\times 10^{-8}\,\mathrm{cm/s}^2.
\end{eqnarray}
\section{\label{Int_accuracy} Sources of computational errors}

Given the very large number of observations for the same spacecraft, the
error contribution from observational noise is very small and not a
meaningful measure of uncertainty. It is therefore necessary to consider
several other effects in order to assign realistic errors. The first
consideration is the statistical and numerical stability of the
calculations. Then there is the cumulative influence of all program modeling errors and data editing decisions. Besides these factors, there are errors that may be attributed to the specific hardware used to run the orbit determination computer codes, together with the algorithms and statistical methods used to obtain the solution. These factors have been
included in the summary of the biases and uncertainties given in
Table~\ref{error_budget}. The largest contributor in this category, and the question we will discuss here, is the reason for and significance of a
periodic term that appears in the data.

\begin{table*}[ht!]
\begin{center}
\caption{Error budget: A summary of biases and uncertainties.
\label{error_budget}} \vskip 20pt
\begin{tabular}{rlll} \hline\hline
Item & Description of error budget constituents & 
Bias~~~~~& Uncertainty \\ & &
$10^{-8}$\,cm/s$^2$~ & $10^{-8}$\,cm/s$^2$~ \\\hline
& & & \\
1. & Sources of acceleration noise external to the craft && \\
& (a) Solar radiation pressure & $+0.03$ & $\pm 0.01$\\
& (b) Solar wind && $ \pm < 10^{-3}$ \\
& (c) Solar corona & & $ \pm 0.02$ \\
& (d) Electromagnetic Lorentz forces && $\pm < 10^{-4}$ \\
& (e) Phase stability and clocks && $\pm <0.001$ \\[10pt]
2. & On-board sources of acceleration noise && \\
& (a) Radio beam reaction force & $+1.10$&$\pm 0.10$ \\
& (b) RTG heat reflected off the craft & $-0.55$&$\pm 0.55$ \\
& (c) Differential emissivity of the RTGs & & $\pm 0.85$ \\
& (d) Non-isotropic radiative cooling of the spacecraft && $\pm 0.48$\\
& (e) Expelled Helium produced within the RTGs 
&$+0.15$ & $\pm 0.16$ \\
& (f) Gas leakage & & $\pm 0.56$ \\
& (g) Variation between spacecraft determinations 
& $+0.17$ & $\pm 0.17$ \\[10pt]
3. & Sources of computational errors && \\
& (a) Accuracy of consistency/model tests & &$\pm0.13$ \\
& (b) Annual term & & $\pm 0.32$ \\
& (c) Other important effects & &$\pm 0.04$ \\[10pt] 
\hline
& Estimate of total bias/error & $+0.90$& $\pm 1.33$ \\
\hline\hline
\end{tabular} 
\end{center}
\end{table*}

{\bf Problem 5.1}. In addition to the constant anomalous acceleration term, an annual sinusoid has been reported.\cite{pioprd} The peaks of the sinusoid occur when the Earth is exactly on the opposite side
of the Sun from the craft, where the Doppler noise is at a maximum. The
amplitude of this oscillatory term by the end of the data interval was 
$\approx0.22 \times 10^{-8}$\,cm/s$^2$. The integral of a sine wave in the acceleration, with angular velocity $\omega$ and amplitude $a_{0},$ yields the following first-order Doppler amplitude, in the fractional
frequency change accumulated over the round-trip travel time of the
radio-signal: 
\begin{equation}
\frac{\Delta \nu}{\nu} = \frac{2a_0}{c \omega}.
\label{lasttwo}
\end{equation}
The resulting Doppler amplitude from the sine wave with amplitude of  
$a_{0} = 0.22 \times 10^{-8}$\,cm/s$^2$ and annual angular velocity 
$\omega= 2 \times 10^{-7}$\,rad/s is $\Delta \nu/\nu = 7.3 \times
10^{-13}$. At the Pioneer downlink S-band carrier frequency of $\approx
2.29$\,GHz, the corresponding Doppler amplitude is 0.002\,Hz, that is,
0.11\,mm/s. 

A four-parameter, nonlinear, weighted, least-squares fit to the annual sine wave was found with the amplitude $v_{\rm at}=(0.1053\pm 0.0107)$\,mm/s, phase $=(-5.3^\circ \pm 7.2^\circ$), angular velocity
$\omega_{\rm at}=(0.0177 \pm 0.0001$)\,rad/day, and bias $=(0.0720 \pm
0.0082$)\,mm/s. Standard data weighting procedures\cite{pioprd} yield post-fit weighted rms residuals of $\sigma_{T}=0.1$\,mm/s (the Doppler error averaged over the data interval $T$). 

The amplitude, $v_{\rm at}$, and angular velocity, $\omega_{\rm at}$, of
the annual term result in a small acceleration amplitude of
$a_{\rm at}=v_{\rm at}\omega_{\rm at} = (0.215 \pm 0.022) \times
10^{-8}$\,cm/s$^2$. The cause is most likely 
due to errors in the navigation programs' determinations of the 
direction of the spacecraft's orbital inclination to the ecliptic. Estimate
the annual contribution to the error budget for $a_P$. 

{\bf Solution 5.1}. First observe that the standard errors for the radial
velocity, $v_r$, and acceleration, $a_r$, are essentially what would be
expected for a linear regression. The caveat is that they are scaled by the root sum of squares of the Doppler error and unmodeled sinusoidal
errors, rather than just the Doppler error. Furthermore, because the error is systematic, it is unrealistic to assume that the errors for $v_r$ and $a_r$ can be reduced by a factor 1/$\sqrt{N}$, where $N$ is the number of data points. Instead, if we average their correlation matrix over the data interval, $T$, we obtain the estimated systematic error of 
\begin{eqnarray}
\sigma_{a_r}^2 = \frac{12}{T^2}\,\sigma_{v_r}^2 =
\frac{12}{T^2} (\sigma_{T}^2 + 
\sigma_{v_{\rm at}}^2),
\label{syserror}
\end{eqnarray}
where $\sigma_{T}=0.1$\,mm/s is the Doppler error averaged over $T$ (not the
standard error of a single Doppler measurement), and $\sigma_{v_{\rm at}}$
is equal to the amplitude of the unmodeled annual sine wave divided by
$\sqrt{2}$. The resulting root sum of squares error in the radial velocity
determination is about 
$\sigma_{v_r}= (\sigma_{T}^2 + \sigma_{v_{\rm at}}^2)^{1/2}=0.15$\,mm/s
for both Pioneer 10 and 11. Our values of $a_P$ were determined over time
intervals of longer than a year. At the same time, to detect an annual
signature in the residuals, we need at least half of the Earth's orbit to be complete. Therefore, with $T = 1/2$ yr, Eq.~(\ref{syserror}) results in an acceleration error of
\begin{equation} 
\sigma_{{\rm a_r}} = \frac{0.50\,{\rm mm/s}}{T}
= 0.32 \times 10^{-8}\,{\mathrm{cm/s}}^2.
\label{aderror}
\end{equation} 
This number is assumed to be the systematic error from the annual term.

\section{\label{budget} Summary of ERRORs AND FINAL RESULT}

The tests discussed in the preceding sections have considered various 
potential sources of systematic error. The results are summarized in
Table~\ref{error_budget}. The only major source of uncertainty not discussed
here is the non-isotropic radiative cooling of the
spacecraft.\cite{pioprd,mpla} Other more minor effects are discussed in
Ref.~\onlinecite{pioprd}. Table~\ref{error_budget} summarizes 
the systematic errors affecting the measured anomalous signal, the
``error budget'' of the anomaly. It is useful for evaluating
the accuracy of the solution for $a_P$ and for guiding possible future
efforts with other spacecraft.

Finally, there is the intractable mathematical problem of how to handle
combined experimental systematic and computational errors. In the end it was decided to treat them all in a least squares uncorrelated
manner.\cite{pioprd} 

{\bf Problem 6.1}. The first column in Table~\ref{error_budget} gives the
bias, $b_P$, and the second gives the uncertainty, $\pm \sigma_P$. The
constituents of the error budget are listed in three categories: 
systematic errors generated external to the spacecraft; on-board
generated systematic errors, and computational errors. Our final result
will be an average
\begin{equation}
a_P = a_{P({\rm exp)}} + b_P \pm \sigma_P,
\label{eq:tot}
\end{equation} 
where
\begin{equation}
a_{P({\rm exp)}} = (7.84\pm 0.01) \times 10^{-8}\,{\mathrm{cm/s}}^2
\label{apexp} 
\end{equation}
is our formal solution for the Pioneer anomaly that was obtained with the available data set.\cite{pioprd} 

Discuss the contributions of various effects to the total error budget and determine the final value for the Pioneer anomaly, $a_P$.

{\bf Solution 6.1}. The least significant factors of our error budget are in the first group of effects, those external to the spacecraft. From
Table~\ref{error_budget} we see that some are near the limit of
contributing, and in total they are insignificant. 

The on-board generated systematics are the largest contributors to the total. All the important constituents are listed in the second group of effects in Table~\ref{error_budget}. Among these effects, the radio beam reaction force produces the largest bias, $1.10\times 10^{-8}$\,cm/s$^2$, and makes the Pioneer effect larger. The largest bias/uncertainty is from RTG heat reflecting off the spacecraft. The effect is as large as $(-0.55 \pm 0.55) \times 10^{-8}$\,cm/s$^2$. Large uncertainties also come from differential emissivity of the RTGs, radiative cooling, and gas leaks, $\pm 0.85$, $\pm 0.48$, and $\pm 0.56$, respectively, $\times 10^{-8}$\,cm/s$^2$. 

The computational errors are listed in the third group of 
Table~\ref{error_budget}. We obtain $b_P \pm \sigma_P = 0.90\times
10^{-8}$\,cm/s$^2$ and 
$\sigma_P = 1.33\times 10^{-8}$\,cm/s$^2$. Therefore, from
Eq.~(\ref{eq:tot}) the final value for $a_P$ is 
\begin{equation}
a_P = (8.74 \pm 1.33) \times 10^{-8}\,\mbox{cm/s}^2.
\label{aP} 
\end{equation}
The effect is clearly significant and remains to be explained. 

\begin{acknowledgments}

The first grateful acknowledgment must go to Catherine Mignard of the
Observatoire de Nice. She not only inquired if a problem set could
be written for the students at the University of Nice, she also made
valuable comments as the work progressed. We also again express our
gratitude to our many colleagues who, over the years, have either collaborated with us on this problem set or given of their wisdom. In this instance we specifically thank Eunice L. Lau, Philip A. Laing, Russell Anania,  Michael Makoid, Jacques Colin, and Fran\c{c}ois Tanguay.
The work of SGT and JDA was carried out at the Jet Propulsion Laboratory,
California Institute of Technology under a contract with the National
Aeronautics and Space Administration. MMN acknowledges support by the U.S.
Department of Energy.

\end{acknowledgments}

\appendix

\section{Allan deviation}
\label{allan_dev}

To achieve the highest accuracy, the Doppler frequency shift of the carrier
wave is referenced to ground-based atomic frequency standards. The
observable is the fractional frequency shift of the stable and coherent
two-way radio signal (Earth-spacecraft-Earth)
\begin{equation}
y(t)=\frac{\nu_{\rm r}(t)-\nu_{\rm t}}{\nu_{\rm t}}
=\frac{2}{c}\frac{dL}{dt},
\label{eq:y}
\end{equation}
where $\nu_{\rm r}$ and $\nu_{\rm t}$ are, respectively, the transmitted and
received frequencies, $t$ is the receiving time, and $2L$ is the overall
optical distance (including diffraction effects) traversed by the photon in
both directions. 

The received signal is compared with the expected measurement noise in $y$,
given by Eq.~(\ref{eq:y}). The quality of such a measurement is
characterized by its Allan deviation, which defined as 
\begin{equation}
\sigma_y(\tau)=\sqrt{\frac{1}{2}\langle(\bar{y}_{i+1}(\tau)
- \bar{y}_{i}(\tau))^2\rangle},
\end{equation}
where ${\bar y}_i= \frac{1}{\tau}\int_{t_i}^{t_i+\tau}y(t)dt$. The Allan
deviation is the most widely used figure of merit for the characterization
of frequency in this context.

For advanced planetary missions such as Cassini, equipped with
multi-frequency links in the X- and Ka-bands, Allan deviations reach the
level of $\sigma_y \approx 10^{-14}$ for averaging times between $10^3$ and
$10^4$\,s. The Pioneer spacecraft were equipped with S-band Doppler
communication systems for which Allan deviations were usually the order ofl
of $1.3 \times 10^{-12}$ and $1.0 \times 10^{-12}$, respectively, for
$10^3$\,s Doppler integration times.


\begin{thebibliography}{99}
\bibitem{foot1} To be precise, the data analyzed in Ref.~\onlinecite{pioprd} was taken between 3 January 1987 and 22 July 1998 for Pioneer 10 (when the craft was 40\,AU to 70.5\,AU distant from the Sun), and that from Pioneer 11 was obtained between 5 January 1987 and 1 October 1990 (22.4 to 31.7\,AU).
\bibitem{batch-seq} To determine the state of the spacecraft at any given time, the Jet Propulsion Laboratory's orbit determination program uses a batch-sequential filtering and smoothing algorithm with process noise.\cite{pioprd} For this algorithm, any small anomalous forces may be treated as stochastic parameters affecting the spacecraft trajectory. As such, these parameters are also responsible for the stochastic noise in the observational data. To better characterize these noise sources, the data interval is split into a number of constant or variable size intervals, called ``batches,'' and assumptions are made for statistical properties of these noise factors. The mean values and the second moments of the unknown parameters are estimated within the batch. In batch sequential filtering, the reference trajectory is updated after each batch to reflect the best estimate of the true trajectory. (For more details consult Ref.~\onlinecite{pioprd} and also references therein.)
\bibitem{bias} We use the symbol $b$ to identify biases in the measured
value of the Pioneer anomaly produced by various sources of acceleration
noise. The uncertainly in the contribution of these noise sources to the
anomalous acceleration is conventionally denoted by $\sigma$. See an example of this usage in Eq.~(\ref{eq:rad-beam}).
\end{thebibliography}
\end{document}